\documentstyle[12 pt,epsfig]{article}
\begin{document}
\title{
\vspace{-2cm} \rightline{\mbox {\normalsize {Lab/UFR-HEP/0109}}}
\vspace{2cm} \bf NC Branes and Hierarchies\\ in Quantum Hall
Fluids}

\author{A.El Rhalami\footnote{arhalami@hotmail.com}, E.M Sahraoui\footnote{sahraoui@theorie.physik.uni-muenchen.de} and E.H.Saidi\footnote{H-saidi@fsr.ac.ma}        \\
\small{Lab/UFR, High Energy Physics, Physics Department, Faculty
of Science.}\\ \small{Av.Ibn Battota, B.P.1014, Rabat, Morocco}}

\maketitle \hoffset=-1cm\textwidth=15cm \vspace*{0.5cm}
\begin{abstract}
We develop the non commutative Chern-Simons gauge model analysis
modeling the description of the hierarchical states of fractional
quantum Hall fluids. For a generic level $n$ of the hierarchy, we
show that the order parameter matrix $K_{ab}$ is given by $
\theta^{-2}Tr(\tau_a \tau_b)$, where $\{\tau_a, 1\leq a \leq n \}$
define a specific set of $n\times n$ matrices depending on the
parameter $\theta$ and the levels $l_a$ of the CS effective field
theory. Our analysis predicts the existence of a third order
tensor of order parameters $C_{abc}$ induced by the external
magnetic field. It is shown that the $C_{abc}$'s are not new order
parameters and are given by $\theta d_{abc}$, where $d_{abc}$ are
numbers depending on the $l_a$'s. We also give the generalized
quantum Hall soliton extending that obtained in the case of the
Laughlin state.
\end{abstract}

\vspace{2cm} \noindent{\bf Key Words}:  Branes physics, NC
Chern-Simons gauge theory, Fractional Quantum Hall fluids,
Hierarchical states and Matrix model.
 \thispagestyle{empty}
\newpage
\setcounter{page}{1}
\section{Introduction}
Recently it has been proposed that non-commutative (NC)
Chern-Simons gauge theory on the $(2+1)$ space may provide a
description of the Laughlin state of Fractional Quantum Hall (FQH)
fluids [1-6]. In this context, it has been shown that the non
commutativity parameter $\theta$ of the Moyal plane is related to
the filling fraction $\nu_L$ of the Laughlin state and so to the
Chern-Simon effective field coupling $\lambda_{CS}$ as $\theta B
\nu_L=\nu_L\lambda_{CS}=1$;  with $B$ the external magnetic field
and the subscript L stands for Laughlin. In [7], see also [8-11],
it has been also conjectured that a specific assembly of a system
of $D0$, $D2$ and $D6$ branes and $F1$ strings, stretching between
$ D2$ and $D6$, has a low energy dynamics similar to the
fundamental state of FQH systems. There, the boundary states of
the F strings ending on the NC D2 brane are interpreted as the FQH
particles. In an external strong magnetic field $B$, represented
by a large number of D0 branes dissolved in D2, the dynamics of
these particles is modeled by a non commutative Chern-Simons
(NCCS) $U(1)$ gauge field theory.\par In this paper we use these
results to develop the NC Chern-Simon theory modeling the
description of the hierarchical states of FQH systems. The point
is that FQH systems with general expressions of the filling
fraction $\nu$ are not all of them of Laughlin type [12]; i.e with
filling fraction $\nu_L=\frac{1}{m}$, $m$ odd integer. Typical
examples are given by states with $\nu(l_1 ,l_2 )=\frac{l_2}{l_1
l_2 -1}$ where $l_1$ and $l_2$ are respectively odd and even
integers. These kind of states are approached using hierarchical
construction ideas [13-25]. In the hydrodynamical approach of FQH
fluids, the $\frac{l_2}{l_1 l_2 -1}$ state can be viewed as
consisting of two components of incompressible fluids; one
describing $\nu_{L,1}= \frac{1}{l_1}$ FQH state while the other
describes the condensation of quasiparticles on the top of the
$\nu_{L,1}= \frac{1}{l_1}$. Put differently, the $\frac{l_2}{l_1
l_2 -1}$ state can be imagined as a composed system of a Laughlin
state of filling fraction $\nu_{L,2}=\frac{1}{l_1 (l_2 l_1 -1)}$
built on an other one with $\nu_{L,1}= \frac{1}{l_1}$ and
satisfying the identity $\nu(l_1 ,l_2 )=\nu_{L,1}+\nu_{L,2} $. As
such the total number of particles can, roughly speaking, be
thought of as given by the sum $N_1 +N_2$, where $N_1$ is
associated with the state of filling fraction $\nu_{L,1}$ and
$N_2$ with the state $\nu_{L,2}$; see ${\it figures}$ 3 and 4 of
section 4. These features apply as well for higher orders of the
hierarchy with $\nu=\nu(l_1,l_2,...,l_n)$. Current examples
correspond to $\nu(3,2)=\frac{2}{5}=\frac{1}{3}+\frac{1}{15}$ and
$\nu(3,2,2)=\frac{3}{7}=\frac{1}{3}+\frac{1}{15}+\frac{1}{35}$.

\par In the field theory approach, the hierarchical states we
refer to here above are described by a $(2+1)$ dimensional system
of coupled CS gauge fields whose action reads for generic levels
$n$ as, [26,27,28]
\begin{equation}
S= \frac{1}{4\pi}\int d^{3}y \
K_{ab}\partial_{\mu}A_{\nu}^{a}A_{\rho}^{b}\epsilon^{\mu\nu\rho}+
J_a^{\mu}A^a_{\mu}.
\end{equation}
In this eq $K_{ab}$ is a $n \times n$ matrix with specific integer
entries defining the order parameters characterizing the fluid and
carrying the interactions between the CS gauge fields.
$J_a^{\mu}$'s are external charge density currents linked for
$n=1$ with the D6 branes charge of the quantum Hall soliton [7,8].
Note that for this case, the above gauge model reduces to the
usual CS effective field theory of the Laughlin state with filling
fraction $\nu_L=K_{11}^{-1}= {1\over m} $. For $n=2$, however, one
deals with a FQH system with filling fraction $\nu(l_1 ,l_2
)=\frac{l_2}{l_1 l_2 -1}$ and so with hierarchical states of level
two. In the Haldane model where $\nu(l_1 ,l_2 )$ may be decomposed
as the sum over $\nu_{L,1}$ and $\nu_{L,2}$, the $K_{ab}$ matrix
reads then as:
\begin{equation}
K_{ab}=\left(\matrix{(2P_1 +1)&-1\cr-1& 2P_{2}\cr}\right)
\end{equation}
where we have set $l_1=2P_1 +1$ and $l_2=2P_2$. Notice that
because of the property $K_{12}\neq 0$; there exists a non trivial
coupling between the two $A^{1}_\mu$ and $A^{2}_\mu$ CS gauge
fields. This feature will play a crucial role in our present study
especially when we build the NCCS theory describing hierarchies by
generalizing Susskind method .
\par Throughout this study we will discuss in details the above
mentioned level two Haldane states by developing an adequate
generalization of the Susskind construction performed for the
Laughlin model. We will also study the case of generic levels of
Haldane other hierarchies. Moreover we build the generalization of
the quantum Hall soliton and give an interpretation of FQH
hierarchies in terms of $Dp$ branes of uncompactified type IIA
string theory.\par The presentation of this paper is as follows.
In section 2 we develop the matrix model used to describe level
two hierarchical states of FQH liquids with a finite number $N$ of
particles. In section 3, we study the NCCS gauge model for the
case of Haldane states at level two and consider the infinite
limit of $N$ where one dimensional matrix fields are mapped to
(2+1)dimensional fields. In this limit the $U(N)$ symmetry is
replaced by $SDiff(R^2)$, the group of area preserving
diffeomorphisms of the $R^2$ plane while the matrix commutator
becomes a Poisson bracket. In section 4 we build the generalized
quantum Hall soliton describing hierarchical states while in
section 5 we give results concerning generic values of the level
$n$ of the hierarchy. Section 6 is deserved for conclusion.
\section{Hierarchy and Matrix Model for FQH states}
In this section we will construct a non-commutative gauge model
for the description of hierarchical states of FQH liquids. This
model is based on an extension of the Susskind analysis made for
the case of the Laughlin state. We first consider the case of a
finite number $N$ of electrons; then we aboard the interesting
limit when $N$ goes to infinity. The determination of the standard
CS effective field theory as the leading term in the NC parameter
$\theta$ will be worked out explicitly. \par To establish the
matrix model for the description of FQH hierarchies, we will adopt
the following strategy: First we consider a toy model, a matter of
introducing
 some general tools
useful for the next steps. Second we develop our matrix model for
the description of hierarchical states of level two for a system
of $N_1 +N_2 =2N$ electrons and finally consider the limit $N$
goes to infinity.
\subsection{General}
To begin consider an electric charged particle, say an electron,
moving in the real plane in presence of an external constant
magnetic field $B$. Classically this particle is parametrized by
its position $x^i (t); \ i=1,2$ and velocity $v^i=\partial_t{
x^i}$. For a $ B$ field strong enough, the dynamics of the
particle is mainly governed by the coupling
\begin{equation}
S[x]=\frac{eB}{2}\int dt{ \varepsilon_{ij} v^i x^j},
\end{equation}
which induces at the quantum level a non commutativity structure
on the real plane; i.e $[x^i,x^j]\propto \epsilon^{ij}/B$. In the
case of $N$ classical particles, without mutual interactions,
parameterized by the coordinates $x^{i}_a (t)$ and velocities
$v^{i}_a$, the dynamics is dominated by  the  $B-x(t)$ couplings
extending eq(3) as $eB\ \varepsilon_{ij}\ \Sigma_{a=1}^N v^{i}_a
x^{j}_a$ and describing a typical strongly correlated system of
electrons showing a quantum Hall effect of filling fraction
$\nu=\frac{N_e}{N_\phi}$; where $N_\phi=\int_{S^2} B$ and $N_e=N$
are respectively the quantum flux and electrons numbers. Quantum
mechanically, there are different field theoretical methods to
approach the quantum states of this system [20], either by using
techniques of non relativistic quantum mechanics [12], methods of
conformal field theory especially for the study of the edge
excitations[14,15] or again by using the CS effective field model
[16] describing the limit $N\rightarrow \infty$ of electrons. In
this case, the CS theory on the $(2+1)$ dimensional space modeling
the FQH Laughlin state of filling fraction $\nu$ is given by the
following action:
\begin{equation}
S[A]=\frac{1}{4\pi \nu}\int d^3y  \epsilon^{\mu\nu\rho}
\partial_{\mu}A_{\nu}A_{\rho}+
\int d^3y J^{\mu}A_{\mu},
\end{equation}
The link between this field action and FQH fluids dynamics has
been studied in details and most of the results in this direction
has been established several years ago [5-18]. However an
interesting observation has been made recently by Susskind [1] and
further considered in [2-6] and [8-11]. The novelties brought by
the study made in [1] is that: (1) Because of the $B$-field, level
$n$  NC Chern-Simons $U(1)$ gauge theory may provide a description
of the Laughlin theory at filling fraction $\nu_L=\frac{1}{m}$. In
this vision, Eq(4) appears then just as the leading term of a more
general theory which reads in general as:
\begin{equation}
S[A]=\frac{1}{4\pi \nu}\int d^3y  \epsilon^{\mu\nu\rho}
[\partial_{\mu}A_{\nu}\ast A_{\rho}+\frac{2i}{3}A_{\mu}\ast
A_{\nu}\ast A_{\rho}]+ \int d^3y J^{\mu}A_{\mu},
\end{equation}
where $\ast$ stands for the usual star operation of non
commutative field theory [29]. (2) The above NCCS $U(1)$ action is
in fact the $N\rightarrow\infty$ of the following matrix model
action:
\begin{equation}
S=\frac{eB}{2}\int dt (
\epsilon_{ij}\sum_{\alpha=1}^{N}[(\dot{X}^i+i[A_{0},X^{i}])X^{j}+
\theta\epsilon^{ij}A_{0}]_{\alpha,\alpha}),
\end{equation}
by substituting the $X^i_{\alpha \beta}$'s as:
\begin{eqnarray}
X^i_{\alpha \beta}&=&y^i \delta_{\alpha \beta}+\theta
\varepsilon^{ij}(A_j)_{\alpha \beta},\\
 \lbrack y^i,y^j\rbrack &=& i\theta \varepsilon^{ij}.
\end{eqnarray}
At this stage let us give some remarks, they concern some
remarkable properties of the above analysis that we will not have
the occasion to address in the present paper: (a) The finite
matrix model (6), which has been conjectured in [17] to describe
fractional quantum Hall droplets, was shown to be equivalent to
the Calogero integrable model [30] providing then another link
between Calogero and Hall systems. (b) The mapping (7), which is
interpreted as describing fluctuations carried by $A_j$ around the
classical solution $X^i=y^i I$, is a kind of background field
splitting. It is formally similar to the gauge splitting one uses
in the derivation of matrix model from the ten dimensional Super
Yang-Mills theory [31] by using dimensional reduction from $(1+9)$
down to $(1+0)$. Note that in the ideal case where $N=1$, eq(6)
reduces to eq(3) with the constraint eq(8); this ideal situation
will be shown later on to be just the leading term of a
hierarchical series; see eq (9) and eq(45). \par The analysis we
gave here above concerns the Laughlin state; the fundamental state
of FQH systems. In what follows we want to generalize it to
include hierarchical states. We will start by considering quantum
states of level two and too particularly those having a filling
fraction $\nu_H (l_1 ,l_2 )=\frac{l_2}{l_1 l_2 -1}$. Later we turn
to the general case.
\subsection{ 1D NC $U(2)$ gauge model}
 In the purpose of studying FQH states with $\nu_H (l_1 ,l_2 )$, let us first
  consider a toy model where the
  $X^{i}(t)=(X^{1},X^{2})$ are two one dimensional
2$\times$2 hermitian matrix fields whose dynamics, in the strong B
regime, is described by the following action:
\begin{equation}
S=\frac{eB}{2}\int dt\ \epsilon_{ij}\ Tr_{u(2)}\ \lbrack(\dot{X}^i
+i[A_{0},X^{i}])X^{j}+\theta\epsilon^{ij}A_{0}\rbrack
\end{equation}
In this eq $A_{0}=A_{0}(t)$ is a one dimensional (1D) gauge field
valued in the $U(2)$ algebra; naively it may be thought of as the
time component of the $U(1)$ CS gauge field to be considered later
on. To fix the idea, the above action may be viewed as associated
with the leading term of a general formula; see eq(20).
\par Since $U(2)= U(1) \times SU(2)$, only the $U(1)$ gauge factor
of the gauge field will contribute in the second term of the
action(9). This action depends linearly on $A_0$ and so it is just
a Lagrange field carrying a field constraint which can be
determined by calculating its equation of motion. In doing so, we
get the following action of the $X^{i}(t)$'s fields
\begin{equation}
S=\frac{eB}{2}\int dt \epsilon_{ij}Tr_{u(2)}(\dot{X}^{i}X^{j}),
\end{equation}
together with the $2 \times 2$ matrix constraint equation,
\begin{equation}
\lbrack X^{i},X^{j}\rbrack = i\theta\epsilon^{ij}I_2.
\end{equation}
Expanding the $X^{i}(t)$ field matrix as follows:
\begin{equation}
X^{i}=X^{i}_{0}I_2 + Z^{i}_{a}\sigma^{a},
\end{equation}
where $ I_2$ stands for the $2 \times 2$ identity matrix and
$\sigma^{a}=(\sigma^{x},\sigma^{y},\sigma^{z})$ are the usual
Pauli matrices. Splitting the $X^{i}_{0}(t)$ field component,
associated with the $U(1)$ factor of $U(2)$, as a sum of constant
$y^{i}$ and a term dependent on time as:
\begin{equation}
X^{i}_{0}(t)=(y^{i} + Z^{i}_{0}(t))I_2,
\end{equation}
 then putting $X^{i}=y^i I_2 + Z^{i}$  back into eq(11), we get the following algebra,
\begin{equation}
\matrix{\lbrack y^{i},y^{j}\rbrack = i \theta\epsilon^{ij}& &
(1)\cr \lbrack y^{i},Z^{j}\rbrack = 0                    & &(2)\cr
\lbrack Z^{i},Z^{j}\rbrack = 0                    & & (3)\cr}
\end{equation}
Eqs (14.2-3) may be further decomposed using the properties of the
Pauli matrices, in particular the Clifford  and the su(2)
algebraic relations. We find,
\begin{eqnarray}
\lbrack y_{i},Z_{a}(t)\rbrack &=& 0\ \ , \ \  a=0,1,2,3. \\
\sum_{a=0}^{3}\lbrack Z_{a}^{i},Z_{a}^{j}\rbrack I_2 +
i\sum_{a,b,c=1}^{3}
\varepsilon^{abc}Z^{i}_{a}Z^{j}_{b}\sigma^{c}&=&0.
\end{eqnarray}
A natural solution of these eqs is obtained by taking $Z_{a}(t)=0,
\ a=0,1,2,4 $, so that eq(14-1) describes just the classical
solution. Therefore the $Z_{a}^{i}(t)$ fields appearing in
eqs(12-13) are interpreted as describing fluctuations around the
classical configuration $X^{i}=y^{i}I_2$. To get the action
describing the $Z_{a}^{i}(t)$ fluctuations, we substitute the
$X^i(t)$'s by the splitting (12) and use eq(14-1), we get
\begin{equation}
S=\frac{eB}{2}\int
dt\epsilon_{ij}Tr_{U(2)}\lbrack(\dot{Z}^{i}+i\lbrack
A_{0},Z^{i}\rbrack )Z^{j}.
\end{equation}
This action is invariant under $U(2)$ automorphisms of the matrix
fields, namely
\begin{eqnarray}
Z^{i\prime}=U^{+}Z^{i}U\nonumber\\
A_{0}^{\prime}=U^{+}A_{0}^{i}U-iU^{+}{\partial\over \partial t}U.
\end{eqnarray}
Eq(17) may  be generalized by including fermions that we have
ignored here above as they are not needed in the present study; it
may also be extended by using, instead of $2\times 2$ matrices,
higher dimensional matrix fields. We will consider this situation
in section 5 when we consider FQH states with filling fraction
$\nu_H (l_1, l_2 ,...,l_n)$.  One of the extensions we are
interested in here, which will be used to describe FQH states at
level two of hierarchy (SL2 for short), is based on taking
hermitian field matrix valued in the $u(2) \oplus u(N)$.
\section{NC Gauge Model for Haldane States}
 To start consider the system of $N_1 +N_2$ electrons on the real plane
parameterized by the coordinates $y^{i}=(y^{1},y^{2})$. $ N_1$
should be thought as the number of electrons associated with the
underlying Laughlin state of filling fraction
$\nu_{L,1}=\frac{1}{l_1}$. $N_2$ is a priori the number of
particles we get after condensation of quasiparticles [5,21]; it
can be thought of as associated with the filling fraction
$\nu_{L,2}= \nu_H (l_1 ,l_2 )-\nu_{L,1}= \frac{1}{l_1 (l_1 l_2
-1)} $. For reasons of simplicity of the formulation of our
effective model, we will suppose that $N_1 =N_2 =N$ and consider
the case of configurations with a finite number $2N$ of electrons
whose coordinates are represented by $2N \times 2N $ dimensional
hermitian matrices $X^{i}(t)$. These are one dimensional fields
valued in the adjoint representation of  $U(2) \oplus U(N)\subset
U(2N)$. Put differently, the $X^{i}$ fields have an expansion
generalizing eq(12) in the sense that each component $Z_a $ is
itself a $N \times N $ hermitian matrix valued adjoint of the
group $U(N)$:
\begin{equation}
Z^{i}_{a}=\sum_{B}T_{B}Z_{a}^{i,B}(t),
\end{equation}
where the $T_B$'s are the $U(N)$ generators. The new matrix model
describing the dynamics of SL2, in presence of a strong B field,
has an action formally similar to eq (6), except now that the
$X^{i}(t)$ and $A_0(t)$ fields are in $Adj_{U(2) \oplus U(N)}$ and
the trace is taken over the states of the  $u(2)\oplus u(N)$
algebra. Thus we have,
\begin{equation}
S=\frac{1}{g^{2}_{2}}\int dt \epsilon_{ij}Tr_{(u(2)\oplus u(N)
)}\lbrack(\dot{X}+i[A_{0},X^{i}])X^{j}+\theta\epsilon^{ij}A_{0}(1+J_{0})\rbrack,
\end{equation}
where $J_0$ is the current density of a given external source and
where $g_{2}$ is a coupling constant to be determined later on; it
carries informations on the SL2 filling fraction $\nu$ and the non
commutativity parameter $\theta$.  The action (20) is symmetric
under the following change extending eq(18)
\begin{eqnarray}
{X}^{i\prime}={\it W}^{+}{X}^{i}{\it W}\nonumber\\
A_{0}^{\prime}={\it W}^{+}A_{0}{\it W}-i{\it W}^{+}{\partial\over
\partial t}{\it W},
\end{eqnarray}
where${\it W}= U\otimes V$, is a unitary transformation of the
$U(2)\otimes U(N)$ gauge group. Setting $U = exp
{(i\sum_{a=1}^3\lambda_{a}\sigma^{a})}$ and $V = exp(i
{\sum_{B=1}^{n^2}\Lambda_{B}T^{B})}$, the infinitesimal form of
the transformations eq(21) reads for the case of $U(N)$ gauge
symmetry for instance as:
\begin{eqnarray}
\delta X^{i}&=&-i\lbrack y^{i},\Lambda\rbrack -i\lbrack
A,X^{i}\rbrack\nonumber\\ \delta A_{0}&=&{\partial\over \partial
t}\Lambda +i \lbrack \Lambda,A_{0}\rbrack
\end{eqnarray}
Before going ahead note that due to eq(14.1), $\lbrack
y^{i},\Lambda\rbrack$ behaves as a derivation since,
\begin{equation}
\lbrack y^{i},\Lambda\rbrack =
i\theta\epsilon^{ij}\partial_{j}\Lambda.
\end{equation}
In the limit $N$ goes to infinity the one dimensional $2N\times
2N$ fields $X^{i}(t)$ and $ A_{0}(t)$ become infinite matrices;
they may be represented by (2+1) dimensional field
$X^{i}(t,y^{1},y^{2})$ and $A_{0}(t,y^{1},y^{2})$ and so is the
$Z^i$ fluctuations around the classical solution; all of them are
valued in the $U(2)$ algebra. For later use let us expand this
field in terms of the $U(2)$ generators as,
\begin{equation}
Z^{i}(y)=Z^{i}_{0}(y)I_2 + \sum_{a=1}^3 Z^{i}_{a}(y)\sigma^{a}.
\end{equation}
 Moreover when $N\longrightarrow\infty$ the $U(2)\otimes
U(N)$ symmetry is also mapped to $U(2)\times SDiff(R^{2})$;
$SDiff(R^{2})$ being the area preserving diffeomorphism group of
$R^{2}$ plane. So the $Z^{i}(t,y^{1},y^{2})$ fluctuations around
the classical solution $y^{i}I$ should be proportional to the
space components $A_{j}(y)$ of the $U(1)$ Chern-Simons gauge field
as
\begin{equation}
Z^{i}(y)\propto\epsilon^{ij}A_{j}(y).
\end{equation}
Since the $Z^{i}(y)$ fluctuations scale as [Z]=[X]=[y]=L while the
gauge field $A_j$ scales as $L^{-1}$, the factor of
proportionality should scale like [X]/[A]=$L^2$ as $\theta$ does.
Here below we study these fluctuations and derive the extension of
eq(7) for the case of SL2.
\subsection{ Generalized Susskind map}
 In the decomposition (24) involving the dimensionless Pauli
 matrices, the scaling behaviour of $X$ is completely carried by the
$Z_{a}^i$ component fields. To convert this expansion in term of
the $A^i$ gauge fields scaling as $(length)^{-1}$,  it is
convenient to introduce a new vector basis
$\{\tau^1,\tau^2,\tau^3\, \tau^4 \} $ of $U(2)$ related to the
standard Pauli matrices basis $\{\sigma^x, \sigma^y, \sigma^z,
\sigma^4 \equiv I_2 \}$ as:
\begin{equation}
\tau^a= \sum_{b=1}^4 \ \Gamma^a_b \ \sigma^{b} \qquad ; \qquad
a,b=1,2,3,4.
\end{equation}
where the entries of the invertible $4 \times 4$ matrix $\Gamma$
scale as $(length)^2$. This change of basis turns out to be very
useful when studying the order parameters of FQH fluids; see eqs
(32) and (34). To fix the ideas, let us give hereafter the
following special choice for $\tau_1$ and $\tau_2$ in terms of
$\sigma^a$'s,
\begin{eqnarray}
\tau_1&=&[\ \alpha_0 I+ \alpha_1 \sigma^x+ \alpha_2 \sigma^y \
],\nonumber \\\tau_2&=&[ \  -\alpha_1 \sigma^x- \alpha_2 \sigma^y+
\alpha_3 \sigma^z \ ],
\end{eqnarray}
where we have set $ \alpha_1=\Gamma^1_1=-\Gamma^2_1, \
\alpha_2=\Gamma^1_2=-\Gamma^2_2, \ \alpha_3=\Gamma^2_3$ and
$\alpha_0=\Gamma^4_4$. Similar formulas may be worked out for
$\tau_3$ and $\tau_4$, but we don't need them for the present
study. We will show later that the $\alpha_a$ parameters in the
above eqs are related to the $l_1$ and $l_2$ order parameters of
the SL2 configurations and to the non commutativity parameter
$\theta$ of the plane. Using this $\tau$ vector basis, we can
expand the gauge fields as
\begin{equation}
A_{i}(y)=\tau_1 A^{1}_{i}(y)+\tau_2 A^{2}_{i}(y)+\tau_3
A^{3}_{i}(y)+\tau_4 A^{4}_{i}(y),
\end{equation}
which upon substituting eq(26), we get the right relation between
the $Z$ and $A$ fluctuations. \par To get the infinitesimal gauge
transformations(22), we have to make use of the correspondence
rules mapping infinite matrices algebra to the space of functions
on the plane. Among them we set: \par (i) In the infinite limit,
$N\times N$ matrix commutators $[F(t),G(t)]_{\alpha \beta}$ are
replaced by the Poisson bracket
$\{F(y),G(y)\}=\varepsilon^{kl}\partial_k F
\partial_l G $. In other words, $$\lim_{N\rightarrow\infty}{i}[F(t),G(t)]\rightarrow
 \theta\ \varepsilon^{kl}\ \ \frac{\partial
 F(t,y)}{\partial{y}^k}\
 \frac{\partial G(t,y)}{\partial{y}^l} .$$ So that the
 infinitesimal gauge transformation reads as,
\begin{equation}
 \delta A_{\mu}= \partial_{\mu}\Lambda +\theta \{ \Lambda,A_{\mu}\}+0(2).
\end{equation}
(ii) the trace $Tr_{U(2)\otimes U(N)}$ operation over the
  $U(2)\otimes
U(N)$ adjoint representation states is mapped for
$N\rightarrow\infty$, to $\int d^{2}y Tr_{U(2)}$; $
lim_{N\rightarrow\infty}Tr_{U(N)}$ should be thought of as $\int
d^{2}y$. Therefore we have, after setting $t=y^0$ and associating
the Dirac delta function $\delta^2(y)$ to the $N\times N$ identity
matrix $I_N$, the following correspondence:
$$\lim_{N\rightarrow\infty}\ \frac{1}{N}\int dt\ Tr_{U(2)\otimes
U(N)}[...](t)\ \rightarrow \ \int d^{3}y \
Tr_{U(2)}[...](y).$$\par Starting from eq(20) and taking
$N\rightarrow \infty$, the one dimensional infinite matrix $X(t)$
is mapped to the $(1+2)$ function $X(t,y)=y +A(t,y)$, where
$A(t,y)$ is a gauge field valued in the $U(2)$ algebra. Putting
back the above expression into eq(20) and following the same
analysis made in [1], we get after replacing $Tr_{U(N)}\rightarrow
\int d^{2}y$, the following NC Chern-Simon gauge theory with a non
abelian $U(2)$ gauge group.
\begin{equation}
S=\frac{1}{4\pi g^2_2}\int d^{3}y\ \ \epsilon^{\mu\nu\rho}\
Tr_{U(2)}\lbrack
\partial_{\mu}A_{\nu}A_{\rho}-\frac{1}{3}A_{\mu}\{A_{\nu},A_{\rho}\}\rbrack
+ {\cal O}(2)
\end{equation}
Like in the Susskind analysis, the ${\cal O}(2)$ terms carry
higher corrections in the NC parameter which can be obtained by
expanding the star product in eq(5). Here we will ignore this
detail and so forget about it. The above action is quite similar
to eq(1) that we are looking for. A careful inspection shows that
eq(30) is not convenient to describe states of level two of the
hierarchy as it contains a non abelian gauge symmetry which is not
allowed for the study of FQH hierarchies. In other words the
expansion(28) and so eq(30) contain too much degrees of freedom,
too much more than those appearing in eq (1). They should be then
reduced down to two gauge fields only.\\

 {\bf Comments} \\

 At first sight, the problem of keeping two gauge fields $B$ and $B^{'}$ amongst the four
  $U(2)$ gauge ones;$${\bf B}=B Y+B^{+}\sigma^{-}+B^{-}\sigma^{+} +B^{'}\sigma^{3},$$
  $Y$ is the generator of the $U(1)$ factor of $U(2)$, is not a major task in gauge theory with matter.
Massless and neutral gauge fields are
   obtained by breaking $U(2)$ gauge group
 down to its Cartan subgroup $U(1)^2$. The two other gauge fields; that is the $B^{\pm}$ fields
 associated with the $U(2)$ step generators $\sigma^{\pm}$, acquire masses
 which by an appropriate
 choice of the Higgs vacuum moduli may be
 supposed to be heavy enough and then integrated out by eliminating $B^{\pm}$ through
 their equations of motion; that is:
  $B^{+}=B^{+}(B,B^{'},B^{-};\partial B,\partial B^{'},\partial B^{-})$ and $B^{-}=B^{-}
  (B,B^{'},B^{+};\partial B,\partial B^{'},\partial B^{+})$.
 As a result one gets an effective field theory depending only on the $B$ and $B^{'}$ massless fields.
  Interactions between $B$ and $B^{'}$, which are generally absent in $U(1) \times U(1)$ theories,
   might come here from the
  substitution of  $B^{+}=B^{+}(B,B^{'},B^{-};\partial B,\partial B^{'},\partial B^{-})$ and
 $B^{-}=B^{-}
  (B,B^{'},B^{+};\partial B,\partial B^{'},\partial B^{+})$; but this issue still needs
  a deeper study.
  In the brane language, this consists to have a large
  splitting of the two D2 branes and so too completely separate world volumes.\par Though
 this  scenario seems to be a natural way to approach hierarchy,
 it is however not the unique one can imagine; an other scenario is to think about
 level two of hierarchy
 as described by bound states of two $D2$ branes with gauge fields such as \begin {eqnarray} C&=& U_{11}B+U_{12}
B^{'} \nonumber \\ C^{'}&=&U_{21}B+U_{22} B^{'} \nonumber,
 \end {eqnarray}
 where $B$ and $B^{'}$ are the gauge fields associated with each of the $D2$
 brane involved in the bound configuration and where $U_{ij}$ are some "similarity"
 transformations. The reason is that, as far as the effective
   field theory eqs(1-2) is concerned, the real symmetric matrix $K_{ab}$ which couples
   the $B$ and $B^{'}$ fields can be diagonalized
   leading to the $C$ and $C^{'}$ eigen functions appearing in the
   above eq.  In this
      scheme, the kinetic terms of the diagonal gauge fields,
      namely  $C\partial{C}+C^{'}\partial{C^{'}}$ contain
      implicitly the $B$-$B^{'}$ interactions as shown here below:
      $$C\partial{C}+C^{'}\partial{C^{'}}= U_{11}^{2}B\partial{B}+
       U_{22}^{2}B^{'}\partial{B^{'}}
       +U_{12}U_{21}\left[B\partial{B^{'}}+B^{'}\partial{B}\right]. $$ In this
 way of doing, interactions between the old gauge fields emerge naturally; but as
  a counterpart a convincing geometrical interpretation is lacking.\par
   To overcome these difficulties, we shall develop here below a phenomenological
method based on imposing adhoc constraints  on the fluctuations
around the classical solution.  We have no rigorous way to derive
them, the unique support we can give now is that they lead to the
right result once used.\\

 {\bf SL2 Constraints} \\

  {\bf (C1)} the gauge fluctuations $A_{j}(y)$ around the
classical solution should be carried by two gauge fields
$A_{j}^{1}(y)$ and $A_{j}^{2}(y)$ instead of the four ones
involved in the $U(2)$ gauge theory. This means that the $U(2)$
gauge field $A_{j}(y)$ should be of the form:
\begin{equation}
A_{j}= \tau_{1}A_{j}^{1} + \tau_{2}A_{j}^{2} \ \ \ \ \ \
\end{equation}
where $\tau_{1}$ and $\tau_{2}$ are as in eqs(27).  \par {\bf
(C2)} In the CS effective model of SL2 eqs (1-2), one sees that
the above mentioned $A_{\mu}^{1}$ and $A_{\mu}^{2}$ gauge fields
are coupled to each other through the $K_{ab}$ matrix. Therefore
we demand that $tr(\tau_a \tau_b )$ has moreover a non diagonal
contribution describing the $A_\mu ^a - A_\mu ^b $ couplings.
\begin{eqnarray}
tr(\tau_{a}^2 )&=& \eta_a  \nonumber \\
 tr(\tau_a \tau_b )&\neq&{  0}, \ \ for\ \
b\neq a. \ \ \ \ \ \
 \end {eqnarray}
In this eq the $\eta_a$ parameter is a numerical constant which,
for the case of Haldane hierarchy, is equal to $ \frac{l_a}{g_2
^2}$ .\\

 {\bf The model for level 2 states}\\

 Putting these two physical
constraints back into the action (30), we get up to the first
order in the non commutativity parameter,
\begin{equation}
S=\frac{1}{4\pi}\int d^{3}y\epsilon^{\mu\nu\rho}\left(
K_{ab}\partial_{\mu}A_{\nu}^{a}A_{\rho}^{b}+C_{abc}A_{\mu}^{a}\{A_{\nu}^{b},A_{\rho}^{c}\}\right)+
{\cal O}(2),
\end{equation}
where $\{A_{\nu}^{b},A_{\rho}^{c}\}$ is the usual Poisson bracket
and where
\begin{eqnarray}
K_{ab} &=& \frac{1}{g^2_2}Tr (\tau_{a}\tau_{b})\ \ \ \ \
(a)\nonumber\\ C_{abc}&=&\frac{1}{g^2_2} Tr(
\tau_{a}\tau_{b}\tau_{c})\ \ \ \ (b)
\end{eqnarray}
The leading term of eq (33) is just the usual CS effective field
model describing the level two hierarchical states of FQH liquids
as shown in eq(1). The second term, however, is the novelty
defining a set of order parameters generated by non-commutativity.
Actually the action (33) can be denoted as $S_2$; it extends the
Susskind action $S_1$ eq(5) for the first order in $\theta$. Both
$S_1$ and $S_2$  may be viewed as the two leading terms of a
hierarchy of functionals $S_n$. Moreover given a hierarchy at the
level two; that is a $2\times 2$ matrix $K_{ab}$, one can compute
the $\alpha_0 , \alpha_1, \alpha_2$  and $\alpha_3$ parameters
involved in eq(27)and then determine the $C_{abc}$ coefficients.
To illustrate the method of work, let us first perform the
calculations for the level two of the Haldane hierarchy.

\subsection{ Haldane Hierarchy}
 The order parameters of the level 2 Haldane state of FQH fluids
are encoded in the $2 \times2$ matrix $K_{ab}$ given by eq(2). So
comparing this equation with eq (34.a), one can compute explicitly
the $\tau_1$ and $\tau_2$ matrices. Straightforward calculations
leads to:
\begin{eqnarray}
g_{2}^{2} &=&2\bar{\alpha}\alpha\nonumber\\
 g_{2}^{2}(2P_{1}+1)&=&2(\alpha_{0}^{2}+\alpha\bar{\alpha})\nonumber\\
2g_{2}^{2}P_{2} &=&2(\alpha_{3}^{2}+\alpha\bar{\alpha}).
\end{eqnarray}
where $P_1$ and $P_2$ are as in eq(2) and where
$\alpha=\alpha_{1}+i\alpha_{2}$. As we see these relations define
actually links between the parameters of the Haldane SL2
configuration and the $\alpha_a$'s. To better see these relations,
let us rewrite eqs(35) into a more convenient form as:
\begin{eqnarray}
2P_{1}&=&  {\alpha_{0}^{2}\over |\alpha|^{2}} \ \ \ \ \in {Z^+}\ \
\ \nonumber\\
 2P_{2}
 &=&1+{\alpha_{3}^{2}\over |\alpha|^{2}}\ \ \ \ \ \in {Z^+}.\ \ \
\end{eqnarray}

Moreover as the $\alpha_a$ moduli scale as $(length)^2$ exactly
like the non commutativity parameter $\theta$ of the Moyal plane
eq(14.1), it is then natural to make the following scaling change,
\begin{eqnarray}
\alpha &=& \theta \eta \nonumber \\ \bar{\alpha }&=& \theta\bar{
\eta},
\end{eqnarray}
where now $\eta$ is a non zero complex dimensionless number. Note
by the way a similar change may be also performed for the
$\alpha_0$ and $\alpha_3$. However and as we will see hereafter,
this feature emerges naturally from the scaling eqs(37). Putting
this change back into these relations, we get on one hand
\begin{equation}
g_{2}=\theta |\eta|\sqrt{2},
\end{equation}
and on the other hand
\begin{eqnarray}
\alpha_{0}&=& \pm\theta|\eta|\sqrt{2p_{1}}\\ \alpha_3
&=&\pm\theta|\eta|\sqrt{2p_{2}-1}.
\end{eqnarray}
Setting $\eta=\eta_1+i\eta_2$ and grouping altogether the above
results, one finds the right fluctuations around the classical
solution $X^i=y^i I $ describing the Haldane $SL2$:

\begin{equation}
X^i=\left(\matrix{X^{i}_{11}& X^{i}_{12} \cr X^{i}_{21}&
X^{i}_{22}\cr}\right),
\end{equation}
where the $X^i _{ab}$'s are given by:
\begin{eqnarray}
X^{i}_{11}&=&y^i+\theta|\eta| \varepsilon^{ij}[(\pm)\sqrt{2p_{1}}
A^{1}_{j}+(\pm)\sqrt{(2p_{2}-1)}A_j^2],\nonumber\\
X^{i}_{12}&=&\theta \eta\varepsilon^{ij}[A_{j}^1-A_{j}^2]\nonumber
\\ X^{i}_{21}&=&\theta \bar{\eta}\varepsilon^{ij}[A_{j}^1-A_{j}^2]
\\ X^{i}_{22}&=&y^i+\theta|\eta| \varepsilon^{ij}[(\pm)\sqrt{2p_{1}}
A^{1}_{j}-(\pm)\sqrt{(2p_{2}-1)}A_j^2],\nonumber
\end{eqnarray}
This is a set of sixteen solutions; they define the generalized
Susskind mapping. Moreover, using eqs (34-b) and (39-40); we can
also compute the cubic coupling $C_{abc}$; we find:
\begin{eqnarray}
C_{111}&=&\pm\theta|\eta|\left[2P_1+3\right]\sqrt{2P_{1}}
\nonumber,\\ C_{112}&=&\pm 2\theta|\eta|\sqrt{2P_{1}}\nonumber,\\
  C_{122}&=&\pm 4\theta|\eta|P_2\sqrt{2P_{1}}
\nonumber,\\
 C_{222}&=&0.
\end{eqnarray}
The remaining other parameters are related to the above ones due
to the cyclic property of the trace. Remark that $C_{abc}$
couplings are indeed proportional to $\theta$ as expected.
\section{Hierarchy and FQH Solitons}
Following [7], see also [8] a fractional quantum Hall phase
similar to the one we have been describing is also observed when
studying the low energy dynamics of brane bounds involving D0, D2
and D6-Branes of the ten dimensional uncompactified type IIA
superstring. Denoting the IIA string coordinate field variables by
$\{t(\tau),\varrho(\tau,\sigma ), \vartheta(\tau,\sigma ) ,
\varphi(\tau,\sigma )
 ,\{y^i(\tau,\sigma )\}_{4\leq i \leq 9; }\}$, $\tau$ and $\sigma$
 are the usual string world sheet variables which should not be confused
 with $U(2)\ \  \sigma^a $ and $\tau^a$ matrices introduced in previous sections,
 the above mentioned D branes bound system, called also quantum Hall
soliton, is built for the case of the Laughlin state as follows:\\

 {\it Quantum Hall Soliton}\\

({\bf a}) One two space dimensional spherical D2 brane
parameterized by $\{ t, \varrho= R ,0\leq \vartheta\leq \pi, 0
\leq \varphi \leq 2\pi, \bf{0_6}\}$. At fixed time, this D2-Brane
is embedded in $R^3 \sim R^+ \times S^2$ and for large values of
the radius, D2 may be thought locally of as $R^{1,2}$ which is
interpreted as the space time of the CS gauge theory.\\

 ({\bf b}) $N$ flat six space dimensional D6 brane parameterized by
$\{t,\bf{0_3}
 ,\{y^i\}_{4\leq i \leq 9}\}$ thought of as an external
 source of charge density $J^0 \propto N\delta^3(x)$ located at the
 origin $(x^1,x^2
,x^3)=(0,0,0)$  of the D2 brane. \\

  ({\bf c}) $N$ fundamental strings F1
 stretching between D2 and D6 and parameterized by $\{t,0 \leq\varrho \leq R,
 \bf{0_2}, \bf{0_6}\}$. The string ends on the D2 brane are associated
 to the electrons of FQH fluids .\\

  ({\bf d}) $M$ D0-branes dissolved into the D2
 brane; They define the flux quanta  associated to the external
 magnetic field $B$ of FQH systems.\par
 In this scheme, the FQH particles in the Laughlin state are described by two $N\times
 N$ hermitian matrices $X^1(t)\sim R\vartheta (t)$ and $X^2 (t) \sim R\varphi
 (t)$ which for large R can approximated by a flat patch of $R^2$.
 In the infinite limit of $N$ and $M$ (strong external magnetic field), the one dimensional matrix
  fields are mapped to the (2+1) fields given by
  $ X^i(t,y)= y^i + \theta\varepsilon^{ij}A_j (t,y)$ as discussed
  in section 3.\par
For states of the level two of hierarchy, and more generally for
generic $n$ levels with $n\geq 2$,  one can also build quantum
solitons by extending the above construction. The key point in the
generation of hierarchical states out of brane bounds is to
suppose that F1 strings end on a collection of $n$ D2 branes in a
specific manner. Let us first present the generalized  quantum
Hall soliton we propose for describing hierarchy; then we make
comments:\\

 {\it Generalized Quantum Hall Soliton}\\

 It is built as follows; see figures 1,2,3 and 4:\\

  ({\bf a}) $n$ coincident spherical D2 brane which we represent as $\{nD2\}$. It has a
   $U(n)$ symmetry group generated by an $n^2$ dimensional basis of $n\times n$
   matrix generators $t^a$.\\

 ({\bf b}) A system containing  (${N_1 +N_2 +...+N_n}$) flat coincident D6 branes to which we refer to as $
 \{{N_a}D6\}_{1\leq a\leq n}$. They are located at the origin of the three space
 (${X^1, X^2, X^3}$) and are associated with the different charge densities
  $J^{0}_a$ one encounters in the effective CS gauge model of FQH fluids; eq(1). \\

  ({\bf c}) A system of $\{{N_1 +N_2 +...+N_n}\}$ fundamental strings F1,
  $ \{{N_a}F1\}_{1\leq a\leq n}$,
 stretching between the $nD2$ and $\{N_a D6\}_{1\leq a\leq n}$. String ends on the $nD2$
  branes are associated with the various kinds of particles involved in the building
   of the hierarchical states  of FQH fluids. The various $N_a$ particles
    are obtained by condensation of quasi-particles which in the
    present language are associated with the $n$  $J^{0}_a$ current
    densities. For simplicity, we suppose that $N_a$'s are all of them equal and so the total number of particles
    is $nN$.\par Moreover these string sets
  are not independent; they do interact as required by the effective CS gauge theory for
  FQH hierarchy. Given two string sets $F1_{(a)}$ and $F1_{(b)}$, their interaction is carried by
  $K_{ab}$ coupling. \\

  ({\bf d}) $M D0$ branes, with, $M=(M_1 +M_2 + ...+ M_n)$,  dissolved in the $nD2$
 branes system. As before, they define the flux quanta associated with the various sets of $F1$
  strings; $ \{{N_a}F1\}_{1\leq a\leq n}$. \\

\begin{figure}[hbt]
\begin{center}
\epsfxsize=10cm \epsffile{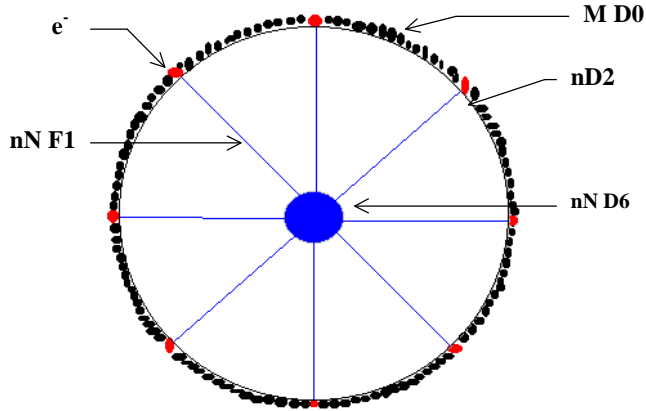}
\end{center}
\caption[]{\it This figure represents a generalized fractional
Quantum Hall Soliton describing the level $n$ of the hierarchy. It
consists of $n$ coincident spherical D2 branes, $nN=\Sigma_{i=1}^n
N_i$ fundamental strings and $M=\Sigma_{i=1}^n M_i$ D0 branes. }
\end{figure}

\begin{figure}[hbt]
\begin{center}
\epsfxsize=8cm \epsffile{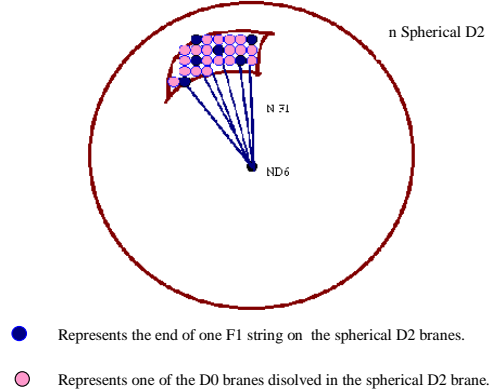}
\end{center}
\caption[empty]{\it represents a portion of the $n$ coincident
spheres describing the generalized fractional Quantum Hall
Soliton.}
\end{figure}

\vspace{2cm} {\it Comments}\\

The generalized quantum Hall soliton describes, amongst others, a
set of $N_1+N_2+...+N_n$ particles. Upon taking all $N_a$'s equal
to $N$, the system has then $nN$ particles and a richer symmetry
which make it more or less simple to handle. We will consider
hereafter this special case.\\
 For finite $N_a$'s, F1 string ends $X^i (t)$ of the full set
$\{NF1\}=\bigcup_{a=1}^n \{{N_a}F1\}$,
  with $nN$ particles, are valued in $Adj_{U(n)\otimes U(N)}$
with very particular coefficients. These coefficients are fixed by
the nature of the $F1_{(a)}$ and $F1_{(b)}$ interactions which,
for the case of Haldane hierarchy, should be in agreement with the
two constraints imposed by the effective CS gauge model of the FQH
systems.\\ String ends of the (${N_a}F1$) subsystem of
$\{({N_a}F1)_{1\leq a\leq n}\}$, are described by the $N\times N$
matrix $X_{a}^i (t)$. This one dimensional field is just the
development of $X^i (t)=\sum_{a=1}^n t^a X_{a}^i (t)$, where the
$n$ $t^a$ hermitian matrices are given by a particular subset of
the $u(n)$ algebra generator basis system. The $X^i (t)$ fields
describe then the full set $\{({N_a}F1)_{1\leq a\leq n}\}$.
Moreover according the analysis of section 3, it is more
convenient to expand $X^i (t)$ as:
\begin{equation}
(X^{i}(t))_{\alpha\beta}=y^{i}\delta_{\alpha\beta} +
\varepsilon^{ij}\ \ \Sigma_{a,b=1}^{n}\ \ [(
A_{j}(t))_{\alpha\beta}^{a}\ \Gamma_{ab}\ (t^b)],
\end{equation}
where $\alpha ,\beta =1,...,N$ and where $\Gamma$ is a  $n\times
n$ matrix whose entries $\Gamma_{ab}$ scale as $(length)^2$.
$\Gamma$ generalizes just the one dimensional non commutative
parameter involved in the analysis on the Laughlin state. \\ In
the limit $N$ infinite, the F1 string ends fill all the space of
 the D2 branes and so
the 1D $N\times N$ matrix $( A_{j})_{\alpha\beta}(t)$ is mapped to
(2+1)D gauge field $ A_{j}(t,y)$. Each set of ($N_a F1$) strings
is then represented by a 2+1 dimensional gauge field $A^{(a)}_\mu$
and consequently the full F1 string ends set $\bigcup_{a=1}^n
\{({N_a}F1)\}$ is described by the gauge field system
$\{A^{(a)}_\mu, 1\leq a\leq n\}$. F1 string interactions are
carried by the $K_{ab}$. Finally note that the D6 branes appear as
external source of charge which couples to the CS gauge fields.
\begin{figure}[hbt]
\begin{center}
\epsfxsize=8cm \epsffile{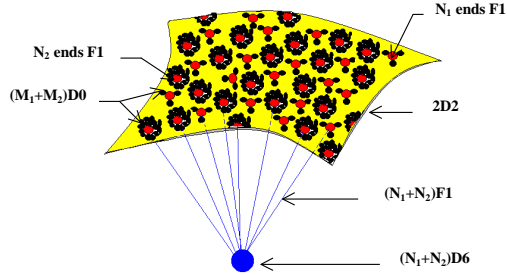}
\end{center}
\caption[empty]{\it  Here is represented the Haldane state of
filling fraction $\nu =\frac{2}{5}$ realized as
$\frac{1}{3}+\frac{1}{15}$. States with 3 D0 branes correspond to
$\nu_L=\frac{1}{3}$ and those with 15 D0 branes for
$\nu_L=\frac{1}{15}$.}
\end{figure}
\begin{figure}[hbt]
\begin{center}
\epsfxsize=8cm \epsffile{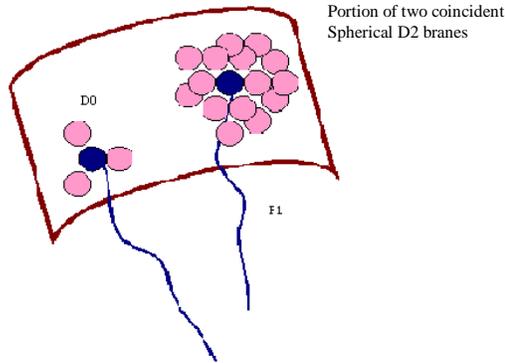}
\end{center}
\caption[empty]{\it Here we show two F1 string ends of the level
two of the hierarchy of filling fraction $\frac{2}{5}$ using the
splitting $\frac{1}{3}+\frac{1}{15}$. The elementary
 $\frac{1}{3}$(resp $\frac{1}{15}$) state is
represented as an F1 string end surrounded by 3 (resp 15) D0
branes.}
\end{figure}
\section{More Results}
Here we give the results for generic levels of the Haldane
hierarchy by following the lines of section 3. In this case FQH
hierarchical states at level $n$ are described, for a finite
number $nN$ of particles, by a one dimensional $nN\times nN$
hermitian matrix $X^i(t)$ field valued in the $u(n)\oplus u(N)$
algebra. The corresponding matrix model for the strong $B$ regime
reads as,
\begin{equation}
S=\frac{1}{g^2_{n}}\int dt \epsilon_{ij}Tr_{u(n)\oplus u(N)
}\lbrack(\dot{X}+i[A_{0},X^{i}])X^{j}+\theta\epsilon^{ij}A_{0}\rbrack,
\end{equation}
where $g_n$ is a coupling constant. This action extends the
Susskind model as well as the formula (20) respectively obtained
by setting $n=1$ and $n=2$.  Eq(45) defines then a sequence of
models in one to one correspondence with FQH hierarchy. For a
finite number $N$, the $ X^i$'s may be treated as
\begin{equation}
X^{i}=[y^{i}+ Z^{i}_{0}]I_{n} + \sum_{a=1}^{n^2 -1}Z^{i}_{a}t^{a},
\end{equation}
where $I_{n}$ is the $n\times n$ identity matrix, the $t^{a}$'s
are the $(n^2 -1) \ su(n)$ generators and where each component
$Z^{i}_{a}$ is itself given by a $N\times N$ matrix of $AdjU(N)$.
As we have noted in section 3, the $Z_{a}^i$ fluctuations around
the classical configuration are not all of them allowed in the
study of FQH hierarchy; only $n$ amongst the $n^2$ ones are
involved in the effective CS model as shown on eq(1). To get the
right fluctuations, we shall follow the method we developed
previously by introducing a new vector basis $\{ \tau^a ; \  1\leq
a\leq n^2
 \}$ of the $u(n)$ algebra. This new basis is related to the old
one as
\begin{equation}
\tau^a=\Gamma^a_b \ t^{b} \ \ ,
\end{equation}
where $\Gamma^a_b$ is an invertible $n\times n$ matrix. Note that
as far as this change is concerned, we will need in practice only
$n$ matrices $\tau^a $, $1\leq a \leq n$ which, without loss of
generality, can be taken as:
\begin{eqnarray}
\tau_1&=&[\gamma_1 I+ (\delta_1 E_1^{+}+ \bar{\delta}_1
E_1^{-})],\nonumber
\\
\tau_a&=&[ -(\beta_a E_{a-1}^{+}+ \bar{\beta}_a E_{a-1}^{-})+
\gamma_a H_{a-1} + (\delta_a E_{a}^{+}+ \bar{\delta}_a
E_{a}^{-})],\ \ 2\leq a\leq n-1 \nonumber,
\\ \tau_n&=&[-(\beta_n E_{n-1}^{+}+ \bar{\beta}_n E_{n-1}^{-})+ \gamma_n
H_{n-1}].
\end{eqnarray}
In this equation the $H_a$'s and $E_a^{\pm}$'s are respectively
the $n\times n$ matrix Cartan generators and Chevalley step
operators of the $su(n)$ algebra while the $\beta_a$'s ,
$\gamma_a$'s and $\delta_a $'s are parameters which should be
related to the order parameters of $SLn$. The $E_a^{\pm}$'s are
the generators associated with the $\alpha_a$ simple roots of the
$su(n)$ algebra. Notice that the above expression for the $\tau_a
$ matrices depend on $5n-4$ real moduli; that is $n$ real
parameters $\gamma_1,\gamma_2,...,\gamma_n$, $n-1$ complex
$\beta_2,...,\beta_n$ and $n-1$ complex
$\delta_1,...,\delta_{n-1}$. These moduli are not all of them
independent; only a subset of them do. Later on we will show that
for the case of Haldane hierarchy there are $n+1$ independent
moduli giving the $n$ CS levels $l_a$ and the non commutativity
parameter $\theta$.
\par In the limit $N$ infinite, the 1D $u(n)\oplus u(N)\ Z^i(t)$
fields are mapped to (2+1)dimensional $ \ Z^{i}(t,y^1,y^2)$ valued
in the $u(n)$ algebra which in turns can be set as
$Z^{i}(y)=\varepsilon^{ij}A_j (y)$ as required by the $U(\infty)
\sim SDiff(R^2)$ invariance. Taking into account all above
features and following the same lines we used for the $SL2$ mode
we get after some algebra,
\begin{equation}
S=\frac{1}{g^2_n}\int d^{3}y\epsilon^{\mu\nu\rho}Tr_{U(n)}\lbrack
\partial_{\mu}A_{\nu}A_{\rho}-\frac{i}{3}A_{\mu}\{A_{\nu},A_{\rho}\}\rbrack
+ 0(2).
\end{equation}
Moreover, as we noted in the case of level two of the hierarchy,
the expansion (49) is not the one needed to describe the FQH
hierarchy;  it involves $n^2$ gauge variables while we need $n$
fields only. This means that eq(46) should be constrained as in eq
(44).\par For the $n-th $ level of the Haldane hierarchy, the SLn
constraint eqs leading to the appropriate result read as:
\begin{eqnarray}
A_{i}(y)&=&\sum_{a=1}^n\tau_a A^{a}_{i}(y),\\ Tr(
\tau_a\tau_{b})-\frac{l_a}{g_n ^2}\  \delta_{ab}&\neq&0 \ \ for\ \
b={a\pm 1},\\ Tr( \tau_a\tau_{b})-\frac{l_a}{g_n ^2}\
\delta_{ab}&=&0 \ \ otherwise.\nonumber
\end{eqnarray}
 The $\tau_a$ solutions of (51) are indeed given by eqs(48). Putting
 these constraint eqs back
into the above action, we get similar relations to those given by
eqs(33); but describe now generic $n-th$ levels of the Haldane
hierarchy.
\begin{eqnarray}
K_{ab} &=& \frac{1}{g^2_n}Tr (\tau_{a}\tau_{b})\ \ \ \ \
(a)\nonumber\\ C_{abc}&=&\frac{1}{g^2_n} Tr(
\tau_{a}\tau_{b}\tau_{c})\ \ \ \ (b)
\end{eqnarray}
These formulas  are then valid for any order $n$ of the hierarchy
and are, in this sense, universal. Furthermore, using the explicit
form of the $K$ matrix of Haldane namely,
\begin{equation}
K_{a b}=l_a \delta_{a, b }-\delta_{a, b -1 }-\delta_{a, b +1 } ,
\end{equation}
we can determine the link between the $\beta_a$ , $\gamma_a$ and
$\delta_a $ parameters appearing in eqs(48) and the $l_a$ order
parameters. We have,
\begin{eqnarray}
g_{n}^{2} &=&(\delta_a\bar{\beta}_a+
\bar{\delta}_a\beta_a)\nonumber,\\
 g_{n}^{2}l_{1}&=&(n\gamma_{1}^{2}+2\delta_1\bar{\delta}_1)\nonumber,\\
g_{n}^{2}l_{a}
&=&2(\gamma_{a}^{2}+\beta_a\bar{\beta}_a+\delta_a\bar{\delta}_a),\\
g_{n}^{2}l_{n} &=&2(\gamma_{n}^{2}+\beta_n\bar{\beta}_n)
\nonumber.
\end{eqnarray}
Since Haldane hierarchical states are specified by the $l_a$
levels of the CS gauge model and the $\theta$ parameter, the
$\beta_a$ and $\gamma_a$ moduli should be constrained. A
convenient choice for the $\beta_a$ and $\gamma_a$ parameters
consists of setting set $\beta_a=\delta_a=\beta$, for all values
of $a$. This permits to have the right degrees of freedom one has
in Haldane theory; that is: $\gamma_1,\gamma_2,...,\gamma_n$ and
$\beta$.  Therefore the $\tau_a$ matrices of eqs (48) are reduced
to:
\begin{eqnarray}
\tau_1&=&{\gamma_1} I+ \beta\ ( E_1^{+}+ E_1^{-}),\nonumber\\
 \tau_a&=& {\gamma_a }H_{a-1}- \beta\ (E_{a-1}^{+}+E_{a-1}^{-}+ E_{a}^{+}+ E_{a}^{-}),
 \ \ 2\leq
a\leq n-1 \nonumber,
\\ \tau_n&=& {\gamma_n}H_{n-1}- \beta\ ( E_{n-1}^{+}+ E_{n-1}^{-}).
\end{eqnarray}
and so eqs(54) is reduced to
\begin{eqnarray}
g_{n}^{2} &=&2\beta^2\nonumber,\\
 g_{n}^{2}l_{1}&=&(n\gamma_{1}^{2}+2\beta^2)\nonumber,\\
g_{n}^{2}l_{a} &=&2(\gamma_{a}^{2}+2\beta^2),\\ g_{n}^{2}l_{n}
&=&2(\gamma_{n}^{2}+\beta^2) \nonumber.
\end{eqnarray}
These eqs may be rewritten in the following equivalent form which
establishes the link between the CS integers $l_1, l_2, l_n$ and
the coupling $g_n $ on one hand and the $\beta$ and $\gamma_a$
moduli on the other hand;
\begin{eqnarray}
l_{1}&=&  (1+{n\gamma_{1}^{2}\over g_{n}^{2}})^\frac{1}{2} \ \ \ \
\ odd \ integer,  \nonumber \\ l_{a}&=& 2 (1+{\gamma_{a}^{2}\over
g_{n}^{2}})^\frac{1}{2} \ \ \ \ \ even \ integer; \ \ 2\leq a\leq
n-1,\nonumber
\\
 l_{n}&=& (1+{2\gamma_{n}^{2}\over g_{n}^{2}})^\frac{1}{2}\ \ \ \ \ even
 \
integer.
\end{eqnarray}
Actually the above eqs constitute a generalization of the Susskind
result on Laughlin state and the level 2 Haldane state we have
obtained in section 3, eqs(38). These analysis correspond just the
two leading modes $SL1$ and $SL2$ of a hierarchy of $SLn$
configurations.
\section{Conclusion}
In this paper we have developed the non commutative Chern-Simons
gauge analysis for the description of the hierarchical states of
fractional quantum Hall liquids. For a generic level $n$ of the
hierarchy, we have shown that  Susskind analysis made for the
Laughlin state is naturally generalization for the hierarchical
one. Using general features on the CS effective field model of FQH
hierarchical states, we have first studied hierarchical states at
level two with a special focus on the Haldane hierarchy and then
considered the generic case. Among our results:
\par (a) The derivation of the matrix model describing a set of a finite number $nN$
of FQH particles which reads as:
\begin{equation}
S=\frac{1}{g^2_{n}}\int dt \ \epsilon_{ij}\ Tr_{u(n)\oplus u(N)
}\lbrack(\dot{X}+i[A_{0},X^{i}])X^{j}+\theta\epsilon^{ij}A_{0}\rbrack,
\end{equation}
where the various quantities appearing in this action were
introduced in the core of this paper. Notice that for $n=1$, this
action coincides with that given in [1] and further elaborated in
[32] in connection with the study of edge excitations.
\par (b) The obtention of the generalized mapping, extending the Susskind change
$X^i=y^i+\theta \varepsilon^{ij}\ A_j(y)$ made for the Laughlin
state, is given by the following $n\times n$ matrix:
\begin{equation}
(X^i_{BD})=\left(\matrix{X^{i}_{11}& X^{i}_{12}&...&X^i_{1n} \cr
X^{i}_{21}& X^{i}_{22}&...&X^i_{2n}\cr .&.&...&.\cr .&.&...&.\cr
X^i_{n1}&X^i_{n2}&...&X^i_{nn}}\right)
\end{equation}
where
\begin{equation}
X^i_{BD}= y^i \delta_{BD}+\theta \varepsilon^{ij}(\sum_{a=1}^n
A^{a}_{j}(y)\ (\tau_a)_{BD} ),
\end{equation}
and where the $\tau_a$'s are as in eqs(55-57). Putting eq(60) back
into eq(58), one gets the well known effective field action (1),
but also corrections induced by $(2+1)$ space-time non
commutativity as shown in eq(49).
\par(c) the proof that the $K_{ab}$ order parameters are indeed related to
 the non commutativity $\theta$ parameter; they are given by $
\theta^{-2}Tr(\tau_a \tau_b)$, where the  \  $\{\tau_a, 1\leq a
\leq n \}$ set is given by a specific system of $n\times n$
matrices depending on $n+1$ moduli $\gamma_1,...,\gamma_n$ in
addition to the coupling constants $g_n$. The $\gamma_a$ moduli
are shown to be related to the $l_a$ integers of the FQH
Chern-Simons effective field theory. These relations were worked
out explicitly for the case of Haldane hierarchy as shown in eqs
(57).
\par (d) our analysis predicts moreover the existence of a tensor
$C_{abc}$ of induced order parameters. This set of order
parameters is shown however to be not a new class as these orders
are not really independent. The $C_{abc}$'s are shown to be given
by $ \theta^{-2} Tr(\tau_a \tau_b \tau_c)\propto \theta d_{abc}$,
where $d_{abc}$ are numbers expressed in terms of the $u(n)$
Chevalley generators $H_a, E_a ^+, E_a ^-$.\par Furthermore, we
have studied the link between Hierarchical states of FQH fluids
and D branes. By extending the construction of refs [2,3]
associated with the Laughlin state, we have built the generalized
quantum Hall soliton supposed to describe generic SLn modes;
$n\geq 2$, as a subsystem. As in the case of the Laughlin state,
the generalized quantum Hall soliton carries here also much more
physics; in particular two coupled CS gauge theories, one
describing the electron fluid and the other the fluid of D0
branes. In the present analysis we have considered the special
situation where all $N_a$'s are equal. We have supposed that the
number $N_a$ of particles one obtains from the $a$-th condensation
is equal to $N$ for any index $a$. The resulting system has a
total number of particles equal to $N_1 +N_2 +...+ N_n = nN$ and a
$U(n)$ symmetry. It would be interesting to explore the general
issue for a system of finite number of particles  where the
$N_a$'s are different and rebuild the underlying effective non
commutative CS gauge model.\\

{\bf Acknowledgements}\\ This project has been supported by the
SARS/99 programme, Rabat University.
\newpage
 \section*{References}
\begin{enumerate}

\item[[1]] \ Leonard Susskind, The Quantum Hall fluid and Noncommutative Chern-Simons Theory. hep-th/0101029.
\item[[2]]\ C. Duval, P. A.Horváthy ,Exotic galilean symmetry in the
non-commutative plane, and the Hall effect, hep-th/0106089
\item[[3]]\ Bogdan Morariu, Alexios
P. Polychronakos, Finite Noncommutative Chern-Simons with a Wilson
Line and the Quantum Hall Effect, hep-th/0106072
\item[[4]]\ Alexios P. Polychronakos,  Quantum Hall states on the cylinder as unitary matrix
Chern-Simons theory, hep-th/0106011
\item[[5]]\ Simeon Hellerman, Mark Van Raamsdonk, Quantum Hall Physics = Noncommutative Field Theory ,hep-th/0103179
         Noncommutative Chern-Simons Solitons
\item[[6]] Dongsu Bak, Sung Ku Kim, Kwang-Sup Soh, Jae Hyung Yee
 Phys.Rev. D64 (2001) 025018

\item[[7]] \ John H. Brodie (SLAC), L. Susskind, N. Toumbas,
How Bob Laughlin Tamed The Giant Graviton From TAUB - NUT SPACE,
    JHEP 0102:003,2001 and hep-th/0010105

\item[[8]]\  Steven S. Gubser, Mukund Rangamani,D-Brane Dynamics and the Qunatum Hall Effect.
                 JHEP 0105:041,2001; hep-th/0012155
\item[[9]]\ Simeon Hellerman, Leonard Susskind; Realizing the Quantum Hall System in String Theory,hep-th/0107200
\item[[10]]\ O. Bergman, J.
Brodie, Y. Okawa;The Stringy Quantum Hall Fluid, hep-th/0107178
\item[[11]]\ Avinash Khare, M. B.Paranjape, JHEP 0104 (2001) 002

\item[[12]]\  R.B.Laughlin, Phys.rev.Lett. 50(1983)1395.\\
E.Fradkin, Field Theories of Condensed Matter Systems, Lecture
Note Series V82(1991)
\item[[13]] \ X.G. Wen and A. Zee ,Phys.Rev.B46(1992)2290.\\ Xiao-Gang Wen, Topological Orders
and Edge Excitations in FQH States. PRINT-95-148
(MIT),cond-mat/9506066.

\item[[14]] \ R.B. Laughlin, Anomalous Quantum Hall Effect: An Incompressible Quantum Fluid
with Fractionally Charged Excitations.
Phys.Rev.Lett.50:1395,1398\\ D.arovas, J.R.Schrieffer,F.Wilczek:
Fractional Statistics in the Quantum Hall
Effect,Phys.Rev.Lett.53(1984)722-723
\item[[15]] \ Safi Bahcall and Leonard Susskind , Fluid Dynamics, Chern-Simons Theory and
 the Quantum Hall Effect; Int.J.Mod.Phys.B5:2735-2750,1991
35
\item[[16]] \ X.G. Wen and A. Zee Phys.Rev.B46:2290-2301,1992; B.Blok and X.G Wen
Phys.Rev.B43:8337,1991, Phys.Rev.B42:8145-8156,1990; ibid
8133-8144,1990.\\
 A.Jellal Int.J.Theor.Phys.37:2751-2755,1998

\item[[17]] \ J. Frohlich and A. Zee, Nucl.Phys.B364:517-540,1991;
Jurg Frohlich, Thomas Kerler  and Emmanuel Thiran; Structure the
set of Incompressible Quantum Hall
Fluids.Nucl.Phys.B453:670-704,1995.
\item[[18]]\ N.Read,Phys.Rev.Lett. 65,(1990)1502.
\item[[19]]\ B.Blok and X.G.Wen, Phys.Rev. B42,(1990)8133-8145.
\item[[20]]\ R.E.Prange and S.M. Girvin, The Quantum Hall effect(Springer, New York,
1987).
\item[[21]]\ F.D.M. Haldane, Fractional Quantization of the Hall Effect:
A Hierarchy of Incompressible Quantum Fluid States.
 Phys.Rev.Lett.51:605-608,1983

\item[[22]]\ Eduardo Fradkin, Lecture Note Series 82 (Field Theories of Condensed matter systems)
1991.
\item[[23]]\ Steven M. Girvin,(Indiana University), "The Quantum Hall Effect: Novel Excitations and Broken Symmetries",cond-mat/9907002.
\item[[24]]\ S.Hellerman and L.Susskind:Realizing the Quantum Hall System in String Theory,
hep-th/0107200.
\item[[25]]\ I. Benkaddour, A. EL Rhalami, E.H. Saidi :Fractional Quantum
Hall Excitations as RdTS Highest Weight State, hep-th/0101188.
\item[[26]]\ H.B.Geyer(Ed):Field Theory, Topology and Condensed matter Physics,Proceedings of the Ninth Chris Engelbrecht Summer School in theoretical Physics. Held at:Storms River Mouth, Tsitsikamma National Parkc, South africa, 17-28 January 1994.Springer 1995. (see) A.Zee,Quatum Hall Fluid
\item[[27]]\ B.Halprin: Statistics of Quasiparticles and the hierarchy of fractional Quantum Hall States, Phys.Rev.Lett.52(1984)1583-1586.
\item[[28]]\ N.Read, Excitation Structure of the Hierarchy Scheme for the fractional Quantum Hall effect, Phys.Rev.Lett.65(1990)1502-1505.
\item[[29]]\ N.Seiberg and E.Witten, JHEP 9909:032,1999, hep-th/9908142
\item[[30]]\ M .Olshanetsky and A.M.Perelomov, Phys.Rept.71(1981)313 and 94(1983)6\\
F.Calogero, J.Math.Phys.12,419(1971)\\
A.P.Polychronakos,Phys.Lett.B 266,29 (1991)
\item[[31]]\ J.Polchinski, String theory, Cambridge Univ Press 1998.
\item[[32]]\ A.P. Polychronakos: Quantum Hall states as matrix Chern-Simons theory, JHEP 0104 (2001) 011,hep-th/0103013.

\end{enumerate}

\end{document}